\documentclass[%
 reprint,
 superscriptaddress,
 amsmath,amssymb,
 aps,
 pra,
]{revtex4-2}

\usepackage{graphicx}
\usepackage{dcolumn}
\usepackage{bm}
\usepackage{ulem}
\usepackage[english]{babel} 

\usepackage{color}
\usepackage{xcolor}
\usepackage{booktabs}
\usepackage{amsmath}
\usepackage{commath}
\usepackage[colorlinks=true,linkcolor=blue,citecolor=blue,urlcolor=blue]{hyperref}

\newcommand{\modified}[1]{\textcolor{black}{#1}}

\begin{document}


\preprint{APS/123-QED}

\title{Cellular Flow Architecture Exposes the Hidden Mechanics of Biological Matter}

\author{Tianxiang Ma}
\thanks{These authors contributed equally to this work}
\affiliation{Niels Bohr Institute, University of Copenhagen, 2100 Copenhagen, Denmark}

\author{Valeriia Grudtsyna}
\thanks{These authors contributed equally to this work}
\affiliation{Niels Bohr Institute, University of Copenhagen, 2100 Copenhagen, Denmark}

\author{Robin V. B\"olsterli}
\affiliation{Niels Bohr Institute, University of Copenhagen, 2100 Copenhagen, Denmark}

\author{Amin Doostmohammadi}
\affiliation{Niels Bohr Institute, University of Copenhagen, 2100 Copenhagen, Denmark}
\affiliation{Corresponding authors: doostmohammadi@nbi.ku.dk}

\begin{abstract}
\noindent Understanding how biomechanical reorganization governs key biological processes, such as morphogenesis and development, requires predictive insights into stress distributions and cellular behavior. While traditional approaches focused on cell motion as a response to stress, we demonstrate that Lagrangian coherent structures (LCSs)---robust attractors and repellers in cellular flows---precede and drive long-term intercellular stress reorganization, physically governed by the mechanical properties of intercellular junctions. We show that this hidden flow skeleton correlates strongly with biomechanical metrics, bridging microscopic cell motion with mesoscopic biomechanics. Specifically, attractors and repellers mark hotspots of compressive and tensile stress enrichment (exceeding tenfold), alongside heterogeneities in cell packing. Notably, these connections remain robust across varying strengths of cell-cell and cell-substrate force transmission. Finally, by linking the attracting regions in the flow skeleton to future cell extrusion spots, we establish a direct link between cell motion and biologically significant outcomes. Together, these findings establish a framework for using cell motion to independently \modified{infer} biomechanical metrics and bridge the scale mismatch between cell motion and biomechanics, potentially offering a new route to interpret mechanosensitive biological processes directly from cell trajectories.
\end{abstract}

\maketitle


Biomechanical reorganizations in cell collectives are central to a wide range of physiological processes, from embryonic development to cancer progression~\cite{valet2022mechanical,wirtz2011physics}. These processes are governed by a suite of biomechanical metrics, including mechanical stress~\cite{ladoux2017mechanobiology}, cell packing~\cite{zehnder2015cell,zehnder2015multicellular}, and cell shape~\cite{Guo2017CellVC}, which collectively influence mechanotransduction and cellular functions. \modified{However, comprehensive quantification of these metrics typically requires distinct measurements and specialized methodologies. In particular, assessing mechanical stress is challenging and often relies on advanced approaches such as force microscopy or fluorescence-based probes~\cite{tambe2011collective,haase2015investigating,ichbiah2023embryo,roffay2024tutorial}.} In contrast, collective cell motion, which is intrinsically linked to the underlying biomechanical state, can be readily quantified~\cite{tambe2011collective,Angelini2011GlasslikeDO,zaritsky2015seeds,bruckner2024learning}. This raises a fundamental question: Can the dynamics of collective cell motion reveal the hidden biomechanical state of a tissue, and if so, how?

Addressing this question requires integrating diverse biomechanical metrics into a cohesive framework, which is both important and challenging. Each metric provides a unique perspective on the mechanical environment of the tissue. Yet, the interplay among these metrics and the dynamics of cell motion, influenced by various factors like substrate stiffness and cell–cell adhesion, generates a complex landscape~\cite{alert2020physical}. Traditional methods often fall short in capturing these interactions comprehensively, necessitating innovative frameworks that can bridge these scales and provide \modified{quantitative} insights.

Traditional approaches to analyzing cell motion rely on Eulerian descriptions, where velocity fields are measured within a fixed laboratory frame using techniques such as particle image velocimetry (PIV)~\cite{thielickePIVlabUserfriendlyAffordable2014} or optical flow~\cite{vig2016quantification}. 
\modified{These methods have revealed important insights, notably the propagating mechanical waves that globally couple tissue kinematics and stress~\cite{serra2012mechanical,notbohm2016cellular,rodriguez2017long,trepat2018mesoscale,boocock2021theory}. However, Eulerian analyses face two inherent challenges. First, they are subject to trajectory mixing: by measuring fields at fixed positions rather than following individual cells, Eulerian methods capture only instantaneous flow information and cannot recover the accumulated deformation history of moving cells. Second, they often suffer from noise and variability at the cellular scale and fluctuate rapidly in time~\cite{flamholz2014quantified,rosen2022mathematical}. Together, these factors hinder the ability to localize when and where mesoscale biomechanical reorganizations emerge. Moreover, Eulerian descriptions are inherently frame-dependent, limiting their ability to capture the intrinsic dynamics of cell collectives~\cite{serra2020dynamic}.} These limitations highlight the need for a frame-independent approach that can bridge the gap between local cell motion and mesoscopic biomechanical reorganizations.


%
\begin{figure}
\centering
\includegraphics[width=1\linewidth]{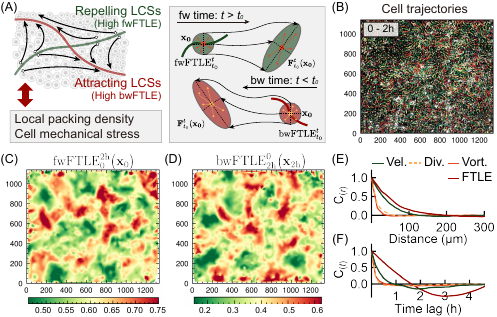}
\caption{\textbf{Lagrangian Coherent Structures (LCSs) in Cell Monolayers.}
\textbf{(A)} Schematic of LCSs in cell collectives and their links to cell mechanical stress and local cell packing density. Green curves indicate LCS repellers (high forward FTLE, fwFTLE) and red curves indicate LCS attractors (high backward FTLE, bwFTLE), with black paths representing cell trajectories. The right panel illustrates fwFTLE as the maximal separation of initially adjacent cells over the interval \([t_0, t]\) (\(t > t_0\)) and bwFTLE as the maximal convergence of initially distant cells over \([t, t_0]\) (\(t < t_0\)). \(F_{t_0}^t(\mathbf{x}_0)\) denotes the Lagrangian flow map, which maps the position \(\mathbf{x}_0\) at time \(t_0\) to its position at time \(t\).
\textbf{(B--D)} Representative cell trajectories overlaid on the bright-field image \textbf{(B)}, \modified{and the corresponding fwFTLE field (\textbf{(C)}, \(\mathrm{fwFTLE}_{0}^{2\mathrm{h}}\bigl(\mathbf{x}_{0}\bigr)\), visualized at the initial tissue configuration \(\mathbf{x}_{0}\)) and bwFTLE field (\textbf{(D)}, \(\mathrm{bwFTLE}^{0}_{2\mathrm{h}}\bigl(\mathbf{x}_{2\mathrm{h}}\bigr)\), visualized at the tissue configuration at 2~h \(\mathbf{x}_{2\mathrm{h}}\)), over a 2~h interval.} FTLE fields are normalized to a scale of 0--1, axes in \(\mu\)m.
\textbf{(E, F)} Spatial \textbf{(E)} and temporal \textbf{(F)} autocorrelation functions comparing FTLE with Eulerian fields—velocity (Vel.), divergence (Div.), and vorticity (Vort.)—demonstrating that FTLE exhibits greater spatiotemporal scales (averaged across 5 independent experiments).}
\label{fig:fig1}
\end{figure}

A promising alternative lies in the Lagrangian framework, which focuses on the trajectories of cells rather than instantaneous velocity fields. Within this framework, Lagrangian coherent structures (LCSs) provide a powerful tool for identifying robust, large-scale patterns in collective cell motion~\cite{serra2020dynamic,haller2015lagrangian}. LCSs act as a ``hidden skeleton'' of the flow, revealing coherent features such as repellers and attractors that persist over time and are computable from sparse and noisy cell trajectories~\cite{mowlavi2022detecting}. These structures are quantified using the finite-time Lyapunov exponent (FTLE), which measures the rate of separation between neighboring trajectories and thus identifies regions of strong repulsion (fwFTLE) or attraction (bwFTLE)~\cite{shadden2005definition}. Prior studies have demonstrated the utility of LCSs in uncovering biological phenomena across scales, from microscopic microtubule self-assembly~\cite{serra2023defect}\modified{, through mesoscopic cell layer collective dynamics~\cite{lee2019dynamics}}, to macroscopic tissue morphogenesis~\cite{serra2020dynamic,mowlavi2022detecting,serra2023mechanochemical}. \modified{Notably, Lee \textit{et al.} related FTLE fields to the leading edge of collective migration~\cite{lee2013quantifying} and to the chaotic dynamics of cell monolayers over long timescales~\cite{lee2016collective}}, while Serra \textit{et al.}~\cite{serra2020dynamic} successfully linked FTLE patterns to large-scale morphogenetic processes during embryonic development. 
\modified{Crucially, however, despite the demonstrated relevance of LCSs to biological processes across scales, their connection to mesoscopic mechanical features}, which is not only experimentally challenging to measure but also constitute core physical mechanisms underlying a broad range of mechanosensitive biological processes---remains largely unexplored.\\

In this study, we demonstrate how this alternative Lagrangian approach can serve as a \modified{quantitative} 
framework for biomechanical reorganizations in cell monolayers. By combining trajectory tracking with Lagrangian Gradient Regression (LGR), we extract robust attractors and repellers of the cellular flow and correlate them with key biomechanical metrics, including intercellular stress and cell packing. We find that the repellers expose regions of elevated tensile stress and reduced local cell packing, while the attractors exhibit the opposite trends. These connections remain robust across variations in substrate stiffness and intercellular adhesion, underscoring its applicability to diverse mechanical contexts. Crucially, we show that the emergence of attractors and repellers precedes biomechanical reorganizations, enabling short-term attractors and repellers to \modified{correlate with}
long-term changes in tissue mechanics. Mechanical perturbations further reveal that the physical mechanism of stress persistence is controlled by cell-cell junctions.
Finally, we demonstrate that cellular flow attractors \modified{are associated with future}
cell extrusion events, linking our framework to a critical mechanosensitive process in tissue homeostasis and disease.\\ 

\noindent {\bf Extracting Lagrangian Coherent Structures (LCSs) from Sparse Cell Trajectories.}
We began by extracting Lagrangian coherent structures (LCSs) within monolayers of Madin-Darby canine kidney (MDCK) cells by computing the finite-time Lyapunov exponent (FTLE) over a time interval \([t_0, t]\) (\hyperref[fig:fig1]{Fig. 1A}), directly from discrete cell trajectories (\hyperref[fig:fig1]{Fig. 1B}). The FTLE is a scalar field that quantifies local flow deformation by identifying regions where initially neighboring particles---cells in this study---either diverge or converge over time, as defined in~\cite{shadden2005definition}:

\begin{equation}
\mathrm{FTLE}_{t_0}^{t}(\mathbf{x}_0) = \frac{1}{|t - t_0|} \log \sqrt{\lambda_{\max}(C_{t_0}^{t}(\mathbf{x}_0))}
\end{equation}

\noindent \modified{where \(C_{t_0}^{t}(\mathbf{x}_0)\) is the right Cauchy–Green strain tensor,} 

\begin{equation}
C_{t_0}^{t}(\mathbf{x}_0) = \bigl(\nabla F_{t_0}^{t}(\mathbf{x}_0)\bigr)^\top \nabla F_{t_0}^{t}(\mathbf{x}_0)
\end{equation}

\noindent \modified{with \(F_{t_0}^{t}(\mathbf{x}_0)\) denoting the Lagrangian flow map that maps the position \(\mathbf{x}_0\) at time \(t_0\) to its position at time \(t\). The largest eigenvalue, \(\lambda_{\max}\), quantifies the maximal stretching of infinitesimally close trajectories, and its logarithm defines the FTLE. Forward FTLE (\(\mathrm{fwFTLE}_{t_0}^{t}\bigl(\mathbf{x}_{0}\bigr)\), with \(t > t_0\)) is computed by tracking cell motion forward in time, thereby quantifying the maximal separation of initially neighboring cells over the interval. Conversely, backward FTLE (\(\mathrm{bwFTLE}_{t_0}^{t}\bigl(\mathbf{x}_{0}\bigr)\), with \(t < t_0\)), is computed by tracking cell motion in the reversed time direction, thus highlighting the maximal convergence of initially distant cells. Together, regions of high fwFTLE and bwFTLE delineate LCS repellers (\hyperref[fig:fig1]{Fig. 1C}) and attractors (\hyperref[fig:fig1]{Fig. 1D}), respectively.}

\modified{Central to this definition is the Jacobian of the flow map, \(\nabla F_{t_0}^{t}(\mathbf{x}_0)\), whose accurate estimation requires sufficiently fine spatial resolution over finite time intervals (see \textbf{Supplementary Methods}). This requirement is particularly challenging in discrete systems such as cell layers, where sampling is sparse and constrained by finite cell size. A conventional approach circumvents this limitation by deriving cell velocities using particle image velocimetry (PIV)~\cite{thielickePIVlabUserfriendlyAffordable2014} and numerically integrating them to generate dense artificial trajectories for FTLE computation. While this procedure improves spatial resolution, it also introduces noise due to oversampling.}

\modified{To overcome these limitations, we implemented the Lagrangian Gradient Regression (LGR) method~\cite{harms2024lagrangian}, which leverages
the principle that, over sufficiently short time intervals, local flow can be approximated as linear. This allows the right Cauchy–Green strain tensor to be calculated incrementally at each step and combined across the total time interval. Specifically, the total time interval \([t_0, t]\) is discretized into \(n\) intermediate steps (\(t_0 < t_1 < \cdots < t_n = t\)), such that the flow map over the entire interval to be expressed as~\cite{brunton2010fast,fenelon2023kinematic}:}
\begin{equation}
\modified{\mathbf{F}_{t_0}^{t_n}(\mathbf{x}_0) = \mathbf{F}_{t_{n-1}}^{t_n}\circ \cdots \circ \mathbf{F}_{t_1}^{t_2}\circ \mathbf{F}_{t_0}^{t_1}(\mathbf{x}_0)}
\end{equation}

\noindent \modified{with the chain rule yielding}
\begin{equation}
\modified{\nabla F_{t_0}^{t_n}(\mathbf{x}_0) = \nabla F_{t_{n-1}}^{t_n}(\mathbf{x}_{t_{n-1}}) \cdots \nabla F_{t_0}^{t_1}(\mathbf{x}_0)}
\end{equation}

\modified{At each short time interval \([t_i, t_{i+1}]\), least-squares regression is applied to estimate \(\nabla F_{t_i}^{t_{i+1}}\). Specifically, a central cell \(\mathbf{x}_{t_i}\) is chosen, and its \(K_n\) nearest neighboring particles \(\mathbf{x}_{j,t_i}\) are identified. Then, we computed the relative displacements \(\Delta \mathbf{x}_{j,t_i} = \mathbf{x}_{j,t_i} - \mathbf{x}_{t_i}\) and compared with their updated values \(\Delta \mathbf{x}_{j,t_{i+1}}\) at the next step. These are then assembled into matrices for regression:}


\begin{equation}
\modified{\begin{aligned}
\mathbf{X}_{t_i} &=
\begin{bmatrix}
\Delta x_{1,t_i} & \Delta x_{2,t_i} & \cdots & \Delta x_{n,t_i}
\end{bmatrix}, \\[6pt]
\mathbf{X}_{t_{i+1}} &=
\begin{bmatrix}
\Delta x_{1,t_{i+1}} & \Delta x_{2,t_{i+1}} & \cdots & \Delta x_{n,t_{i+1}}
\end{bmatrix}
\end{aligned}}
\end{equation}

\modified{The deformed positions \(\mathbf{X}_{t_{i+1}}\) are related to the initial positions \(\mathbf{X}_{t_i}\) through a linear mapping \(\mathbf{A}\):}

\begin{equation}
\modified{\mathbf{X}_{t_{i+1}} = \mathbf{A}\,\mathbf{X}_{t_i}}
\end{equation}

\noindent \modified{where the optimal \(\mathbf{A}\), approximating \(\nabla F_{t_i}^{t_{i+1}}\), minimizes:}

\begin{equation}
\modified{\mathbf{A} = \arg\min_{\mathbf{A}} 
\left( 
\frac{1}{2} \left\| \mathbf{K}^{\frac{1}{2}} (\mathbf{X}_{t_{i+1}} - \mathbf{A}\mathbf{X}_{t_i}) \right\|_F^2 + \frac{\gamma}{2} \left\| \mathbf{A} \right\|_F^2 \right)}
\end{equation}

\noindent \modified{Here, \(\|\cdot\|_F\) denotes the Frobenius norm, the weighting matrix \(\mathbf{K}\) is set to the identity to avoid introducing additional hyperparameters, and \(\gamma\) is a small regularization parameter (\(\gamma \ll r\), where \(r\) is the mean intercellular distance). Unless otherwise specified, the number of nearest neighbors (\(K_n\)) was set to 40, corresponding to a regression radius of \(80\,\mu\mathrm{m}\), which is approximately the velocity correlation length (see \hyperref[fig:fig1]{Fig.~1E}). Further details of the regression parameters are provided in the \textbf{Supplementary Methods}.}

\modified{Validation of LGR against the conventional PIV-based FTLE method~\cite{onu2015lcs} is provided in the \textbf{Supplementary Methods} and \textbf{Supplementary Fig.~S1}. Briefly, both methods yielded consistent FTLE patterns, but the PIV-based method introduced noise due to oversampling, whereas LGR produced smoother, more continuous fields.}

Finally, we analyzed the spatiotemporal autocorrelations of FTLE fields and compared them with instantaneous Eulerian metrics (velocity, divergence, and vorticity). FTLE fields exhibited stronger spatial and temporal correlations (\hyperref[fig:fig1]{Fig. 1E,F}), with slower decay rates that effectively filter out short-term oscillations irrelevant to large-scale biomechanical reorganizations~\cite{trepat2009physical,rodriguez2017long,zehnder2015cell,zehnder2015multicellular}. This highlights LCS as a hidden skeleton of the cellular flow and a powerful tool for linking cell motion to mesoscale biomechanical processes.\\

\noindent {\bf Attractors and Repellers Mark Hotspots for Mechanical Stress Enrichment in Cell Monolayers.}
\begin{figure*}[t!]
\centering
\includegraphics[width=\linewidth]{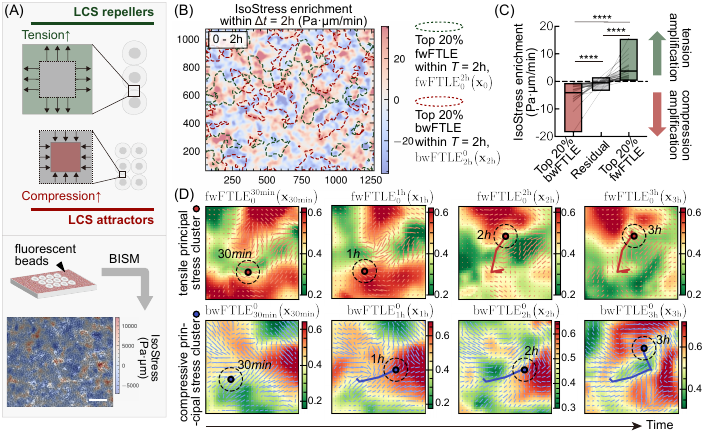}
\caption{\textbf{Lagrangian Coherent Structures (LCSs) Track Intercellular Stress Amplifications.}
\textbf{(A)} Schematic linking LCS attractors and repellers to isotropic intercellular stress. Bottom: Representative isotropic intercellular stress map from Bayesian Inversion Stress Microscopy (BISM) overlaid on a bright-field image (scale bar: 200 \(\mu\)m).
\textbf{(B)} Representative map of isotropic intercellular stress (IsoStress) enrichment over \(\Delta t = 2\)~h, overlaid with regions of the top 20\% high bwFTLE (red dashed lines) and high fwFTLE (green dashed lines) within 0--2~h time interval (\(T = 2\)~h).
\textbf{(C)} Quantification of isotropic intercellular stress enrichment, showing significant compressive amplification in high bwFTLE regions and tensile amplification in high fwFTLE regions. Data from 5 independent experiments span various FTLE intervals (\(T\)) and stress enrichment windows (\(T=\Delta t = \)30~min, 1~h, 2~h, 3~h, 4~h, and 5~h). Each gray line represents an individual sample/time interval; black lines denote the mean. \modified{Statistical significance was assessed using repeated-measures one-way ANOVA with Tukey’s multiple comparisons test (****\(p < 0.0001\)).}
\textbf{(D)} Tensile principal (top) and compressive (bottom) stress clusters tracked over (\(T=\) 30~min, 1~h, 2~h, and 3~h, \modified{overlaid with the corresponding FTLE fields: \(\mathrm{fwFTLE}_{0}^{T}(\mathbf{x}_T)\) (fwFTLE transported along trajectories to \(\mathbf{x}_T\)) and \(\mathrm{bwFTLE}_{T}^{0}(\mathbf{x}_T)\).}
Clusters (red: tensile, blue: compressive) are identified based on director alignment within the black dashed circle. Motion trajectories of clusters are shown as colored lines, and principal stresses are visualized as light-colored directors. FTLE fields are displayed as underlying colormaps (normalized to a scale of 0--1).}
\label{fig:fig2}
\end{figure*}
%
To establish a robust link between cell migration and mechanical stress, we first explored the quantitative relationships between LCS repellers/attractors and intercellular stress (\hyperref[fig:fig2]{Fig. 2A}). Time-lapse bright-field images of MDCK cell monolayers were acquired for LCS calculations, with cell trajectories extracted using a self-trained Cellpose model~\cite{stringer2021cellpose} and Trackmate algorithms~\cite{trackmate2017} (see \textbf{Supplementary Methods} for details). We computed backward and forward \( \text{FTLE}_{t_0}^{t} \) fields over time intervals \( T = |t - t_0| \) of 30~min, 1~h, 2~h, 3~h, 4~h, and 5~h based on these trajectories (see evolutions of bwFTLE and fwFTLE fields in \textbf{Supplementary Movie~S1}). LCS attractors and repellers were defined as regions with the top 20\% bwFTLE and top 20\% fwFTLE values, respectively; regions exhibiting both high bwFTLE and high fwFTLE were excluded, as they correspond to Lagrangian saddles~\cite{rockwood2018tracking} and do not represent purely attracting or repelling dynamics. In parallel, we measured intercellular stress using Bayesian Inversion Stress Microscopy (BISM)~\cite{tambe2011collective,balasubramaniam2021investigating}, which infers intercellular stress based on the force equilibrium between cell–substrate traction and intercellular forces (\hyperref[fig:fig2]{Fig. 2A}).

We focused on the isotropic component of the stress tensor, \((\sigma_{xx} + \sigma_{yy})/2\), since it plays a key role in modulating mechanobiological functions~\cite{Saw2017TopologicalDI,guillamat2022integer,balasubramaniam2025dynamic}. Here, positive values denote tensile stress and negative values indicate compressive stress. The evolution of isotropic intercellular stress, together with corresponding live-cell imaging, is presented in \textbf{Supplementary Movie~S2}. Considering that LCSs integrate cell motion over defined time intervals, we hypothesized that the stress most directly corresponding to LCS is its enrichment---quantified as the time rate of stress change within a time window \(\Delta t\).

To test this, we overlaid the LCS attractors and repellers with the stress enrichment map. \hyperref[fig:fig2]{Fig. 2B} illustrates an example for the 0–2~h interval (\(T=\Delta t=2\)~h): regions corresponding to the top 20\% high \(\text{bwFTLE}_{2\mathrm{h}}^{0}\) (LCS attractors, outlined by red dashed lines) exhibit enrichment in compressive stress, whereas regions corresponding to the top 20\% high \(\text{fwFTLE}_{0}^{2\mathrm{h}}\) (LCS repellers, outlined by green dashed lines) are enriched in tensile stress. The evolution of stress enrichment, along with the corresponding LCS attractor and repeller formations, is visualized in \textbf{Supplementary Movie~S3}. Quantitative analysis across multiple time scales (\(T =\Delta t=\) 30~min, 1~h, 2~h, 3~h, 4~h, and 5~h; \hyperref[fig:fig2]{Fig. 2C}) confirmed the significance of this trend. Remarkably, although averaged across distinct time intervals from short to long term, the enrichment magnitudes of compression (3.91~Pa·µm/min) and tension (3.94~Pa·µm/min) in attracting and repelling regions, respectively, exhibit over a 10-fold amplification compared to those measured in the residual regions (0.38~Pa·µm/min). To further ensure the robustness of these observations, we varied the threshold for defining high FTLE regions from the top 30\% to the top 10\%. The trends observed with the top 20\% threshold were consistently reproduced across all thresholds, as demonstrated in \textbf{Supplementary Fig. S2}.

For comparison, we repeated the analysis using instantaneous Eulerian metrics (\textbf{Supplementary Fig. S3}), including velocity magnitude, divergence, and vorticity. Regions corresponding to the top/bottom 20\% of these fields were assessed for their relationships with stress enrichment. In contrast to the strong correlations observed with LCSs, only subtle relationships were detected using these instantaneous metrics. Furthermore, to assess the robustness of the LCS predictions on intercellular stress, we applied mechanical perturbations by knocking down E-cadherin (E-cad) and altering substrate stiffness to modify cellular interactions~\cite{CHARRAS2018R445} (see \textbf{Supplementary Methods} for details). Notably, the correlations between LCS attractors/repellers and the enrichment of compressive/tensile stress persisted under these perturbations (\textbf{Supplementary Fig. S4}).

To better illustrate the spatiotemporal dynamics of stress reorganization relative to LCS evolution, we applied a clustering-based visualization approach. Specifically, we identified regions exhibiting coordinated tensile or compressive principal stress using a spatial alignment metric and tracked clusters that persisted for at least 3~h~\cite{zaritsky2014propagating,zaritsky2015seeds} (see \textbf{Supplementary Methods} and \textbf{Supplementary Fig. S5}). \modified{We then overlaid the positions of stress clusters at time \(T\) with the FTLE fields visualized at the corresponding tissue configuration \(\mathbf{x}_T\); that is, \(\mathrm{fwFTLE}_{0}^{T}(\mathbf{x}_T)\) (forward FTLE transported along trajectories from \(\mathbf{x}_0\) to \(\mathbf{x}_T\), see \textbf{Supplementary Fig. S6A}) and \(\mathrm{bwFTLE}_{T}^{0}(\mathbf{x}_T)\).}
This visualization reveals that tensile stress clusters tend to move alongside regions of elevated fwFTLE (see top row of \hyperref[fig:fig2]{Fig. 2D}; each panel displays the tracked tensile stress clusters at \(T=\) 30~min, 1~h, 2~h, and 3~h, respectively, overlaid with the corresponding fwFTLE field), while compressive stress clusters co-move with regions of high bwFTLE (\hyperref[fig:fig2]{Fig. 2D}, bottom row). 

\modified{To further quantify the cluster-tracking visualizations, we performed a complementary quantitative analysis that calculates stress changes along individual cell trajectories. Specifically, we classified cells into the top 20\% fwFTLE group (\textbf{Supplementary Fig.~S6A}), the top 20\% bwFTLE group (\textbf{Supplementary Fig.~S6B}), and the residual population. We then quantified local stress enrichment by comparing neighborhood stress values as cells move from \(\mathbf{x}_0\) to \(\mathbf{x}_T\), using the same spatial range applied in the FTLE regression calculations. This confirms that cells experiencing local neighboring attraction are enriched in compressive stress, whereas cells experiencing local neighboring repulsion are enriched in tensile stress (see \textbf{Supplementary Fig.~S6C}).}

Collectively, these findings demonstrate that cells within LCS attractors exhibit enriched compressive stress, whereas those within LCS repellers display enriched tensile stress. Importantly, these correlations persist across various time scales and under distinct mechanical perturbations.

\modified{Building on these correlations, we sought to move beyond correlation and preliminarily assess the predictive potential of LCSs. To this end, we implemented a proof-of-concept machine learning framework inspired by Schmitt et al.~\cite{schmitt2024machine}, originally developed to predict traction forces from focal adhesion images. In our adaptation, forward and backward FTLE fields were used as inputs to predict the spatial distribution of isotropic stress enrichment (\textbf{Supplementary Fig. S7A}). Despite limited data, the predicted maps closely matched experimental measurements (\textbf{Supplementary Fig. S7B-D}), supporting the feasibility of learning-based prediction of intercellular stress enrichment from Lagrangian features.}\\

\noindent \modified{{\bf LCSs Encode Both Isotropic and Anisotropic Stress Patterns.}}
\modified{Recognizing that Lagrangian Coherent Structures (LCSs) capture both isotropic and anisotropic modes of deformation in cellular flows~\cite{serra2020dynamic}, we extended our analysis to examine the relationship between FTLE fields and intercellular anisotropic stress transmission \cite{tambe2011collective}. Specifically, we computed the maximum shear stress, defined as \((\sigma_{1} - \sigma_{2})/2\), where \(\sigma_{1}\) and \(\sigma_{2}\) are the maximum and minimum principal stresses, respectively (representative maps are shown in \textbf{Supplementary Fig.~S8A,B}).}

\modified{Following the same approach used for FTLE–isotropic stress analysis, we overlaid LCS attractors and repellers with the anisotropic stress enrichment maps (\textbf{Supplementary Fig.~S8C}). As shown in \textbf{Supplementary Fig.~S8D}, high bwFTLE regions exhibited reduced maximum shear stress, whereas high fwFTLE regions showed increased shear stress. 
Although the spatial distribution of maximum shear stress is noisier and more heterogeneous than that of isotropic stress~\cite{tambe2011collective}, the observed trends are statistically significant when considering all tested samples and time intervals.}

\modified{While these findings highlight the dual sensitivity of FTLE to both stress components, our primary focus remains on its relationship with isotropic stress. To confirm that this relationship arises specifically from isotropic modes of deformation---and not merely from mixed or anisotropic effects---we isolated the isotropic component of Lagrangian deformation by computing \cite{santhosh2025coherent}:}
\begin{equation}
\text{iso} \, \Lambda_{t_0}^t(\mathbf{x}_0) = \frac{1}{|t - t_0|} \log \left|\det\left(\nabla F_{t_0}^{t}(\mathbf{x}_0)\right)\right|
\end{equation}
\modified{where \(\left|\det\left(\nabla F_{t_0}^{t}(\mathbf{x}_0)\right)\right|\) captures local isotropic expansion or contraction. As shown in \textbf{Supplementary Fig.~S9}, regions with high forward \(\text{iso} \, \Lambda_{0}^{T}\) exhibit enrichment in tensile stress, while those with high backward \(\text{iso} \, \Lambda_{T}^{0}\) exhibit enrichment in compressive stress. These trends closely mirror those observed for FTLE (\hyperref[fig:fig2]{Fig. 2C}), providing mechanistic validation that the FTLE–isotropic stress coupling originates from underlying isotropic deformation.}\\

\noindent {\bf LCS Attractors and Repellers Mark Heterogeneities in Local Cell Packing.}
\begin{figure}[t!]
\centering
\includegraphics[width=\columnwidth]{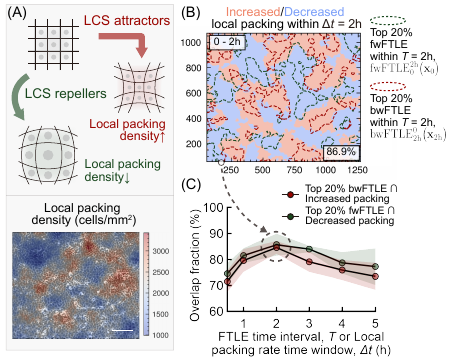}
\caption{\textbf{Interplay between Lagrangian Coherent Structures (LCSs) and Changes in Local Cell Packing.}
\textbf{(A)} Schematic illustrating the changes in local packing within regions of attracting LCSs (high bwFTLE) and repelling LCSs (high fwFTLE). Bottom panel: Representative maps of local packing density overlaid on the bright-field image (scale bar: 200 \(\mu\)m).
\textbf{(B)} Example spatial overlap between LCS attractors (red dashed: high bwFTLE)/repellers (green dashed: high fwFTLE) and areas where the local packing rate is above (orange) or below (blue) the mean over 0–2 h (overlap fraction for this example is 86.9\%).
\textbf{(C)} Overlap fractions between LCS attractors/repellers and areas of increased (or decreased) local packing across time intervals (\(T=\Delta t\) = 30 min, 1 h, 2 h, 3 h, 4 h, and 5 h; mean ± s.d., 5 independent experiments). The gray circle indicates the example data shown in \textbf{(B)}.}
\label{fig:fig3}
\end{figure}
Key biomechanical metrics, including cell packing density and cell size, play critical roles in regulating mechanosensitive cellular functions~\cite{Guo2017CellVC,Neurohr2020RelevanceAR}. \modified{These metrics are not only influenced by cellular stress~\cite{pajic2024epithelial}, for example through fluid exchange between cells~\cite{mcevoy2020gap,yang2022shaping}, but are also tightly coupled to collective kinematics. We therefore hypothesized that Lagrangian coherent structures (LCSs) can also predict reorganizations in cell packing and size. Given the inverse relationship between packing and size, we focused our analysis on the link between LCSs and cell packing (\hyperref[fig:fig3]{Fig. 3A}).}
Local packing density was calculated from Cellpose segmentations (see \textbf{Supplementary Methods} for details). Its temporal evolution, together with corresponding time-lapse bright-field imaging, is presented in \textbf{Supplementary Movie~S4}.

Following the approach used for intercellular stress, we overlaid LCS attractors (the top 20\% bwFTLE regions) and repellers (the top 20\% fwFTLE regions) onto the time rate maps of local packing density. As shown in \textbf{Supplementary Fig. S10}, regions of attracting LCSs exhibit an increase in local packing density, whereas LCS repellers display the opposite trends. Building on these, we masked regions exhibiting increased or decreased local packing density using a threshold based on the average time rate and quantified the spatial overlap between these regions and the LCS attractors/repellers using overlap fraction, defined as \( \frac{S_{\text{FTLE}} \cap S_{\Delta \rho}}{S_{\text{FTLE}}} \). Here, \( S_{\text{FTLE}} \) represents the area of the high bwFTLE (or fwFTLE) regions, and \( S_{\Delta \rho} \) denotes the area of increased (or decreased) local packing density. \hyperref[fig:fig3]{Fig. 3B} exemplifies this spatial correlation, with an overlap fraction of 86.9\% showing a clear correspondence between high \(\text{bwFTLE}_{2\mathrm{h}}^{0}\) regions (red dashed lines) and regions of increased packing (orange masks), as well as between high \(\text{fwFTLE}_{0}^{2\mathrm{h}}\) regions (green dashed lines) and regions of decreased packing (blue masks). As shown in \hyperref[fig:fig3]{Fig. 3C}, this overlap remains robust across all time scales, further reinforcing the relationship between LCSs and local packing density. 
\modified{The observed reduction in overlap at extended intervals is likely attributable to the loss of cell tracks over longer time windows, which diminishes the ability to accurately compute FTLE fields from increasingly sparse trajectory data. By contrast, the reduced overlap at short intervals ($<2$~h) likely reflects intrinsic fluctuations in cell packing driven by intercellular fluid exchange~\cite{zehnder2015multicellular}, consistent with prior studies reporting a minimum in cell density temporal autocorrelation at 2 h~\cite{zehnder2015cell}.} 
The evolution of the local packing density, along with the corresponding LCS attractor/repeller formations, is visualized in \textbf{Supplementary Movie~S5}. 
\modified{As with isotropic stress, we repeated the analysis using instantaneous Eulerian metrics, including velocity magnitude, divergence, and vorticity. These exhibited weaker and less persistent overlap with packing changes compared to FTLE (\textbf{Supplementary Fig.~S11}), highlighting the distinct implications of the Lagrangian framework.}
Finally, we confirmed that these patterns persist under mechanical perturbations through E-cad KO and substrate stiffness alterations (\textbf{Supplementary Fig. S12}).

Taken together, these results underscore the unique capability of LCSs to \modified{quantify}
the spatiotemporal evolution of local cell packing.\\

\begin{figure*}[ht]
\centering
\includegraphics[width=1\linewidth]{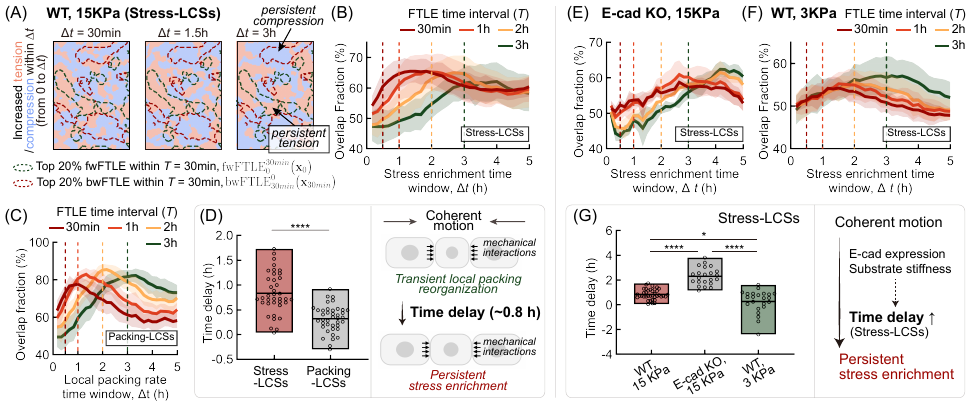}
\caption{
\textbf{Short-Term Lagrangian Coherent Structures (LCSs) \modified{Correlate with} Persistent Enrichment of Intercellular Stress.}  
\textbf{(A)} Overlap between LCS repellers (top 20\% \(\text{fwFTLE}_{0}^{30\,\text{min}}\), green dashed lines) and LCS attractors (top 20\% \(\text{bwFTLE}_{30\,\text{min}}^{0}\), red dashed lines) with regions of enriched tensile (orange masks) and compressive (blue masks) stress. Stress enrichment is computed over time windows of 30~min, 1.5~h, and 3~h, while the FTLE interval remains fixed at 30~min.  
\textbf{(B)} Quantification of the overlap fraction between LCS repellers/attractors and regions of enriched tensile/compressive stress. The stress enrichment time window varies in 10-min increments, with curves shown for FTLE intervals of 30~min, 1~h, 2~h, and 3~h (mean ± s.d., 5 independent experiments).
\textbf{(C)} Similar analysis as in \textbf{(B)}, but for overlap between LCS repellers/attractors and local packing density changes, showing a minor time delay between the FTLE interval and the time window during which local packing density changes (decrease/increase) show the strongest overlap with LCS repellers/attractors.
\textbf{(D)} Left: Time delay between the FTLE interval and the time window during which biomechanical changes (stress and local cell packing) show the highest overlap with LCS repellers/attractors in wild-type cells on 15~kPa substrates (WT, 15~kPa; mean ± s.d., n = 5, with measurements for fwFTLE and bwFTLE intervals of 30~min, 1~h, 2~h, and 3~h). \modified{Statistical significance was assessed using Welch's two-tailed t-test (****\(p < 0.0001\)).} Right: Schematic illustrating the sequence of coherent motion, cell packing density reorganization, and intercellular stress enrichment. 
\textbf{(E)} Overlap fractions in E-cadherin knockout MDCK cells on 15~kPa substrates (E-cad KO, 15~kPa) between LCS repellers and enriched tensile stress, and between LCS attractors and enriched compressive stress, for FTLE intervals of 30~min, 1~h, 2~h, and 3~h (mean ± s.d., n = 3). Results indicate enhanced stress enrichment persistence compared to WT cells on the same 15~kPa substrate.  
\textbf{(F)} Overlap fractions in WT cells on 3~kPa substrates (WT, 3~kPa) between LCS repellers and enriched tensile stress, and between LCS attractors and enriched compressive stress, for FTLE intervals of 30~min, 1~h, 2~h, and 3~h (mean ± s.d., n = 3). Results indicate reduced stress enrichment persistence compared to WT cells on the stiffer 15~kPa substrate. 
\textbf{(G)} Time delay between the FTLE interval and the window during which enriched tensile and compressive stress regions exhibit the highest overlap with LCS repellers and attractors is compared across conditions (WT, 15 kPa; E-cad KO, 15 kPa; WT, 3 kPa). Data are shown as mean ± s.d. (n = 3 for E-cad KO, 15 kPa and WT, 3 kPa; n = 5 for WT, 15 kPa), with measurements for fwFTLE and bwFTLE intervals of 30~min, 1~h, 2~h, and 3~h. \modified{Statistical significance was assessed using Welch's one-way ANOVA with Dunnett's T3 multiple comparisons test (*\(p<0.05\), ****\(p<0.0001\)).} Right: Schematic illustrating the influence of E-cadherin expression and substrate stiffness on the time delay between FTLE patterns and persistent stress enrichment.
}
\label{fig:fig4}
\end{figure*}
%

\noindent {\modified{\bf Short-Term LCSs Correlate with Long-Term Persistent Changes in Intercellular Stress.}}
Thus far, our analysis has focused on correlations between LCSs and biomechanical metrics within the same time windows. However, a critical question remains: \modified{do short-term LCS dynamics relate to longer-term biomechanical patterns?}
To address this, we expanded our analysis to determine whether correlations between LCSs and biomechanics depend on matching the FTLE time interval (\(T\)) with the time windows (\(\Delta t\)) used for calculating biomechanical changes.
To address this, we fixed the FTLE time interval and computed its spatial overlap with biomechanical changes while continuously varying the time window used for calculating biomechanical changes in 10-min increments. \modified{Specifically, both time windows were aligned at their start; for example, FTLE over \(0{-}30~\text{min}\) was compared with stress enrichment over \(0{-}30~\text{min}\), \(0{-}40~\text{min}\), …, \(0{-}5~\text{h}\).}
Remarkably, our results revealed that LCSs \modified{correlate with}
intercellular stress enrichment that persists beyond their computed FTLE time intervals, with peak overlaps occurring after the FTLE interval.

\hyperref[fig:fig4]{Fig. 4A} illustrates this phenomenon, showing the overlap between high \(\text{fwFTLE}_{0}^{30\text{min}}\) (green dashed lines) and high \(\text{bwFTLE}_{30\text{min}}^{0}\) (red dashed lines) with regions of tensile enrichment (orange masks) and compressive enrichment (blue masks). Here, the overlap fraction further increases when comparing the FTLE field formed after 30~min with stress enrichment calculated at later time points. This overlap persisted for stress reorganization measured between 1.5 and 2~h before decaying, yet remained well above 50\% even at 3~h.

\begin{figure*}[ht]
\centering
\includegraphics[width=1\linewidth]{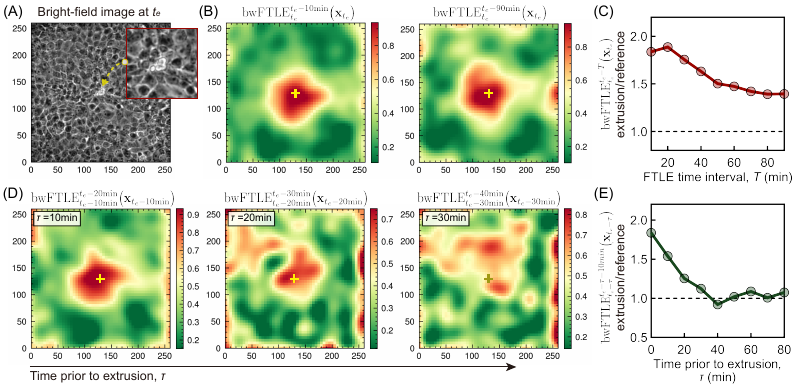}
\caption{\textbf{Lagrangian Coherent Structures (LCSs) \modified{Associate with} Cell Extrusion Events.}
\textbf{(A)} Representative bright-field image of a cell extrusion event at time \(t_e\) (axes in \(\mu\)m). The inset shows a zoomed-in view of the extrusion event centered in the image. 
\textbf{(B)} Left: bwFTLE fields (normalized to a scale of 0--1) calculated from cell trajectories between \(t_e - 10\) min and \(t_e\), denoted as \(\text{bwFTLE}_{t_e}^{t_e-10\,\text{min}}\). Right: bwFTLE fields (normalized to a scale of 0--1) calculated from cell trajectories between \(t_e - 90\) min and \(t_e\), denoted as \(\text{bwFTLE}_{t_e}^{t_e-90\,\text{min}}\). The bwFTLE fields are averaged over 75 extrusion events, with the extrusion sites aligned at the center of the field of view, marked by a yellow cross.
\textbf{(C)} Relative bwFTLE at extrusion sites, defined as the ratio of the averaged bwFTLE at the extrusion sites to that of other regions, across different time intervals (\(T\)). Each scatter point represents the relative \(\text{bwFTLE}_{t_e}^{t_e-T}\), calculated using cell trajectories from \(t_e - T\) to \(t_e\), where \(T\) spans 10 min (corresponding to the left panel of \textbf{(B)}) to 90 min (corresponding to the right panel of \textbf{(B)}) in 10-min increments. Results are averaged over 75 extrusion events.
\textbf{(D)} bwFTLE fields (normalized to a scale of 0--1) calculated using 10-min time interval at specific times \(\tau\) prior to extrusion. Left: bwFTLE calculated from trajectories between \(t_e - 20\) min and \(t_e - 10\) min (\(\tau = 10\) min). Middle: bwFTLE calculated from trajectories between \(t_e - 30\) min and \(t_e - 20\) min (\(\tau = 20\) min). Right: bwFTLE calculated from trajectories between \(t_e - 40\) min and \(t_e - 30\) min (\(\tau = 30\) min). The bwFTLE fields are averaged over 75 extrusion events, with extrusion sites centered and marked by a yellow cross.
\textbf{(E)} Quantitative results showing the relative bwFTLE at extrusion sites, calculated using 10-min trajectories at various times \(\tau\) prior to extrusion.}
\label{fig:fig5}
\end{figure*}

\hyperref[fig:fig4]{Fig. 4B} further validates this persistent correlation across FTLE time intervals of 1~h, 2~h, and 3~h. In contrast, no persistence was observed for local cell packing, where overlaps with high FTLE regions were confined to the corresponding FTLE time intervals (see \hyperref[fig:fig4]{Fig. 4C} and \textbf{Supplementary Fig. S13}). We also quantified the time delay between the FTLE time interval and the time window during which biomechanical changes exhibit peak overlap with LCS repellers/attractors (\textit{i.e.} high fwFTLE/bwFTLE regions). 
As shown in \hyperref[fig:fig4]{Fig. 4D}, this time delay is significantly greater for stress compared to density, suggesting that attracting and repelling regions exhibit persistent amplifications of compressive and tensile stress, respectively. These results indicate that reorganization of LCS patterns precedes mechanical reorganization, suggesting that coherent cellular motion actively shapes mechanical outcomes. 
\modified{In contrast, the reorganization of cell packing, which is more tightly coupled to the underlying cell trajectories than stress, occurs concurrently with cellular flow.}

Because intercellular stress transmission is largely regulated through adherens junctions~\cite{bazellieres2015control,yang2022local}, we hypothesized the delayed stress realignment following coherent cell motion depends on the mechanical properties of cell-cell junctions. To test this, we disrupted intercellular junctions by knocking out E-cadherin (E-cad KO) and then fixed the FTLE time interval to calculate its spatial overlap with intercellular stress enrichment across varying time windows. Interestingly, E-cad KO cells exhibited an extended delay in stress realignment compared to wild-type cells (\hyperref[fig:fig4]{Fig. 4E}), reflecting an increased time lag between coherent cell motion and stress enrichment. Conversely, wild-type cells cultured on soft substrates, characterized by enhanced E-cad expression and stronger intercellular junctions (\modified{\textbf{Supplementary Fig. S14}})~\cite{vazquez2022effect, balasubramaniam2021investigating, wang2016substrate}, displayed a more synchronous overlap between FTLE patterns and stress changes (\hyperref[fig:fig4]{Fig. 4F}). The differences were further quantified in \hyperref[fig:fig4]{Fig. 4G}, where E-cad KO cells on 15~kPa substrates show significantly higher delays, and wild-type cells on 3~kPa substrate show significantly lower delays compared to wild-type cells on 15~kPa substrates. This is in line with the recent notion that stronger intercellular junctions can more efficiently transmit stress to neighboring cells~\cite{schoenit2024force}, whereas weaker junctions impair long-range stress transmission, leading to prolonged local stress enrichment.

Collectively, these results demonstrate that \modified{LCSs calculated from short-term cell trajectories are correlated with long-term reorganizations in intercellular stress.}
Moreover, the persistence of stress changes following coherent motion depends on the mechanical properties of cell-cell junctions: systems with stronger junctions exhibit more synchronous overlaps between FTLE patterns and stress changes, whereas systems with weaker junctions show delayed overlaps. Importantly, we also confirmed that these perturbations have minimal effects on the synchronous correlations between LCSs and changes in cell packing density (see \textbf{Supplementary Fig. S15}), underscoring the specificity of LCSs in independently disentangling distinct biomechanical factors.\\

\noindent {\modified{\bf Correlation of LCSs with Mechanosensitive Biological Processes: Cell Extrusion Sites.}}
Beyond biomechanical metrics, we explored the potential of LCSs to \modified{describe} mechanosensitive biological processes.
To illustrate this capability, we investigated cell extrusion---a cell elimination mechanism essential for maintaining homeostasis that regulated by both mechanical stress and cell packing~\cite{Saw2017TopologicalDI,balasubramaniam2025dynamic,kocgozlu2016epithelial,fadul2018forces}. Here, 75 extrusion sites were analyzed, each cropped to a 260\,$\mu$m square with the extrusion event centered (see \hyperref[fig:fig5]{Fig. 5A}).

First, we examined whether attractor/repeller regions correlate with extrusion events. To this end, we calculated bwFTLE and fwFTLE fields using cell trajectories before and after the extrusion time \(t_e\), with time intervals (\(T\)) ranging from 10 to 90 min. Averaged over the 75 events, the bwFTLE fields calculated from trajectories prior to extrusion (\(\text{bwFTLE}_{t_e}^{t_e-T}\)) consistently displayed significantly elevated values at the extrusion sites, regardless of whether a short-term (\(T=10\) min; see left panel in \hyperref[fig:fig5]{Fig. 5B}) or long-term (\(T=90\) min; see right panel in \hyperref[fig:fig5]{Fig. 5B}) time interval was used. To quantify this trend, we calculated the relative bwFTLE by dividing the average bwFTLE within a circular region (with a radius equal to the average cell radius) centered at the extrusion site by the average bwFTLE outside that region. For all time intervals from 10 to 90 min, the relative bwFTLE at the extrusion sites exceeded 1. \modified{We also confirmed significantly elevated bwFTLE at extrusion sites compared to surrounding regions (\textbf{Supplementary Fig. S16}).} Conversely, following extrusion, fwFTLE values were elevated at these sites \modified{(see relative fwFTLE in \textbf{Supplementary Fig. S17} and corresponding statistical comparisons in \textbf{Supplementary Fig. S18})}. These demonstrate that cell extrusion events are associated with LCS attractors and repellers. \modified{In addition, this correlation was robust to variation in the regression neighborhood size (\textit{i.e.}, the number of nearest neighbors used in the FTLE calculation), as confirmed by tests with neighborhood sizes ranging from 5 to 60 (20 neighbors were used in \hyperref[fig:fig5]{Fig.~5}). At the same time, increasing the neighborhood size---shifting the analysis from localized motion toward broader patterns---gradually reduced both the magnitude and statistical significance of elevated bwFTLE at extrusion sites (\textbf{Supplementary Fig.~S19}), consistent with the localized nature of extrusion events \cite{Saw2017TopologicalDI}.}

Building on our earlier finding that short-term LCSs \modified{are linked to}
long-term stress enrichment, we further assessed whether LCSs \modified{associate with future extrusion events}.
In this analysis, we computed bwFTLE using cell trajectories ending at \(t_e - \tau\) (instead of at \(t_e\) as in \hyperref[fig:fig5]{Fig. 5B,C}), where \(\tau\) denotes the time before extrusion. \hyperref[fig:fig5]{Fig. 5D} shows bwFTLE fields computed from 10-min intervals (\(T=10\) min) ending at \(t_e - \tau\) for \(\tau = 10\), 20, and 30 min. Interestingly, these fields exhibited elevated values at the future extrusion sites, although the magnitude of the elevation decreased with increasing \(\tau\) (\hyperref[fig:fig5]{Fig. 5E}). \modified{Quantitative analysis confirmed that trajectories ending 20 min prior to extrusion show significantly higher bwFTLE at extrusion sites compared to surrounding regions (\textbf{Supplementary Fig. S20}).}

These findings demonstrate that attracting LCSs, as indicated by high bwFTLE values, \modified{serve as short-term markers that correlate with future extrusion events.}
The shorter \modified{lead}
window for extrusion relative to long-term stress enrichment likely reflects the multifactorial nature of extrusion, which is modulated not only by the persistent enrichment of mechanical stress~\cite{balasubramaniam2025dynamic} but also by transient changes of local cell packing~\cite{kocgozlu2016epithelial,fadul2018forces}, which we have shown to be synchronously linked with LCSs reorganizations with small temporal lag.\\

\noindent {\bf Discussion.}
Our findings reveal how the hidden architecture of underlying cellular flows, captured by Lagrangian coherent structures (LCSs), are intimately linked with key biomechanical changes and can drive long-term reorganization of intercellular stress. By establishing a unified framework that relies solely on cell trajectories to extract robust LCS attractors and repellers from the cellular flow field, we first bridge the scale mismatch between microscopic cell motion and mesoscopic biomechanics. We then demonstrate strong correlations between LCSs and crucial biomechanical metrics. Specifically, LCS attractors and repellers mark hotspots characterized by significant enrichments of compressive and tensile stress—exceeding 10-fold amplifications, respectively. Moreover, LCS attractors accurately capture heterogeneous regions of elevated local cell packing whereas LCS repellers identify the opposite. Notably, these correlations persist across diverse mechanical environments, including substrates of varying stiffness and conditions of disrupted cell-cell adhesion via E-cadherin knockout (E-cad KO).

Most importantly, we discover that LCSs not only correlate with intercellular stress reorganizations occurring during the corresponding cell motion, \modified{but also exhibit alignments with}
long-term stress enrichment over extended time windows. For example, LCS attractors and repellers computed from short-term cell trajectories (30~min) \modified{are associated with}
enriched compressive and tensile stress lasting for at least 5~h (\hyperref[fig:fig4]{Fig. 4B}). Comparisons across different E-cad expression levels and substrate stiffnesses indicate that the persistence of stress changes following coherent motion is modulated by the mechanical properties of cell-cell adhesions: systems with stronger junctions exhibit more synchronous overlaps
between LCS patterns and stress changes, whereas systems with weaker junctions display delayed stress realignment. Building on this feature, we further illustrate \modified{that LCS attractors correlate with future cell extrusion events}
---a pivotal mechanosensitive process in tissue homeostasis and disease. These findings underscore the potential of our approach to \modified{reveal quantitative connections with}
a wide range of biologically significant mechanosensitive events~\cite{granero2024nucleocytoplasmic}, solely from cell motion data.

\modified{While our experiments directly probed the role of cell–cell adhesion, the observed delays and synchrony in stress realignment can also be understood in the broader context of adhesion crosstalk. Intercellular stress transmission is mediated by adherens junctions~\cite{bazellieres2015control,yang2022local}, whose effectiveness is modulated by integrin–cadherin crosstalk with focal adhesions~\cite{mui2016mechanical}. Consistent with this, E-cadherin knockout cells---with weaker junctions and compensatory strengthening of focal adhesions~\cite{schoenit2024force}---exhibited prolonged delays in stress realignment, whereas wild-type cells on soft substrates---with enhanced junctional adhesion (see \textbf{Supplementary Fig. S14}) and weakened focal adhesions~\cite{de2017single}---showed more synchronous stress transmission. This interpretation supports the view that stronger intercellular junctions, together with weaker focal adhesions, promote collective stress propagation~\cite{schoenit2024force}, whereas weaker junctions combined with enhanced focal adhesions bias cells toward more individual behavior and impaired cell-cell stress transmission~\cite{mertz2013cadherin,balasubramaniam2021investigating}. Thus, although our data specifically address cell–cell adhesion, they are consistent with a broader principle whereby persistent stress enrichment reflects the balance between intercellular and cell–substrate adhesion in tissues.}

\modified{These novel insights from Lagrangian analyses emerge from two complementary perspectives. First, fundamentally, LCSs provide a Lagrangian measure that follows individual cells over time, accumulating their deformation history and revealing frame-invariant attracting/repelling patterns often invisible to instantaneous Eulerian velocity, vorticity, or divergence fields. This advantage is evidenced by our results showing that Eulerian metrics struggle to identify and spatially map regions where stress changes concentrate (\textbf{Supplementary Fig. S3}). Second, by using LCSs to filter noise and extract coherent patterns in cell groups, our framework identifies mesoscopic features rather than detailed local fluctuations. 
Notably, Eulerian studies---exemplified by mechanical wave analyses~\cite{serra2012mechanical,notbohm2016cellular,boocock2021theory}---investigate large-scale links between cell motion and stress, having established global relationships between these fields. This provides the essential mechanistic foundation for the correlations we identify using Lagrangian methods. However, our framework focuses on a complementary perspective by spatially localizing coherent motion patterns and revealing where stress reorganization concentrates over finite timescales. This feature has potential to be extended toward predictive stress mapping, as preliminarily demonstrated in our proof-of-concept machine learning framework (\textbf{Supplementary Fig. S7}).}

\modified{In addition to the potential applications for identifying spatiotemporal relationships with biological functionalities,}
another interesting avenue for future research with our proposed framework is exploring the potential relationships between LCSs and active nematic behaviors in cell collectives. Recent studies using microtubule–kinesin mixtures have shown that the motion of active nematics is controlled by the dynamics of attracting and repelling LCSs, whose motion is in turn mediated by the formation of topological defects~\cite{serra2023defect}. Considering that active nematic behaviors in cell collectives have been extensively identified in recent years~\cite{Saw2017TopologicalDI,kawaguchi2017topological,duclos2018spontaneous,guillamat2022integer,doostmohammadi2022physics}, it would be interesting to explore the potential links between the cell nematic behaviors, like nematic ordering and defects dynamic, to the coherent motions in cell layers.

From the methodological perspective, the major strength of the LCSs framework lies in its solely reliance on kinematic information. In this study, we utilized cell trajectories, which, by considering the sparse nature of cell collectives, reduced noise and enhanced the accuracy of our analysis. Additionally, LCSs can be identified from continuous velocity fields by artificially integrating the velocity to reconstruct trajectories, as demonstrated in our FTLE calculations using PIV-derived velocity fields (see \textbf{Supplementary Fig. S1A,B}). This adaptability makes the framework broadly applicable across biological systems, including those where individual trajectories are unavailable and only velocity field measurements are feasible, such as chromatin flows~\cite{shaban2020hi} and bacterial biofilms~\cite{grobas2021swarming}. 

Moreover, the applicability of LCSs extends beyond motion data; it can be employed for analyzing other vector fields to filter out localized and transient transport effects, thereby identifying robust structures. For instance, analyzing the traction force fields with LCSs could identify stable force topologies, while employing it on cell-polarity vector fields~\cite{jain2020role} may reveal persistent polarization regions. These applications present a promising approach for bridging the gap between various microscopic properties and mesoscopic and macroscopic biological processes. Importantly, the correlations between LCSs and mechanical stress and packing density identified in this study may not be restricted to biological systems. Recent findings indicate similar correlations between LCSs and polymeric stress fields in long-chain polymers~\cite{kumar2023lagrangian}, suggesting that these relationships could be universal across time-dependent dynamical systems. Future studies could further enhance the predictive power of LCSs by machine learning approaches~\cite{schmitt2024machine,bruckner2024learning}, \modified{extending beyond the stress prediction we demonstrate here to include prediction of events such as cell extrusion}, broadening its applications in both biological and non-biological contexts.

\section*{Acknowledgments}
A. D. acknowledges funding from the Novo Nordisk Foundation (grant no. NNF18SA0035142 and NERD grant no. NNF21OC0068687), Villum Fonden (grant no. 29476), and the European Union (ERC, PhysCoMeT, 101041418). Views and opinions expressed are however those of the authors only and do not necessarily reflect those of the European Union or the European Research Council. Neither the European Union nor the granting authority can be held responsible for them.

\section*{Author contributions}
A.D. designed the project. T.M. implemented and analyzed the Lagrangian coherent structures computations. V.G. performed the experiments and analyses. R.B. contributed the computational tools. A.D. supervised the project. T.M., and A.D. wrote the manuscript, with input from all authors.

\section*{Data and materials availability}
Code has been deposited in GitLab at \url{https://gitlab.nbi.ku.dk/active-intelligent-matter/lcs-within-cell-collectives.git}. The raw live-cell imaging, cell tracking, cell local packing, cell size, and Bayesian Inversion Stress Microscopy data supporting the findings of this study are available at \url{https://sid.erda.dk/sharelink/fLFtnEsdze}.

\bibliography{zHenriquesLab-Mendeley}

\end{document}